\documentclass[reprint, aps, prb,preprintnumbers, floatfix, longbibliography]{revtex4-2}
\usepackage[utf8]{inputenc}
\usepackage{amsmath}
\usepackage{amssymb}
\usepackage{verbatim}
\usepackage{subcaption}
\usepackage{graphicx}
\usepackage{float}
\usepackage{booktabs}
\usepackage{wrapfig}
\usepackage{xcolor}
\usepackage{mathtools}
\usepackage{hyperref}

\definecolor{colorblue}{RGB}{4,4,236}
\hypersetup{
    colorlinks=true,  
    linkcolor=colorblue,   
    citecolor=colorblue,    
    urlcolor=colorblue, 
    filecolor=colorblue   
}

\begin{document}  

\preprint{APS/123-QED}

\title{Single-shot near-field reconstruction of metamaterial dispersion}

\author{Eugene~A.~Koreshin$^1$}
\email{evgeniy.koreshin@metalab.ifmo.ru}

\author{Denis~Sakhno$^{1,2}$}


\author{Jim~A.~Enriquez$^1$}

\author{Pavel~A.~Belov$^{1,2}$}

\affiliation{$^1$School of Physics and Engineering, ITMO University, Kronverksky Pr. 49, 197101 St. Petersburg, Russia}

\affiliation{
$^2$School of Engineering, New Uzbekistan University, Movarounnahr str. 1, 100000, Tashkent, Uzbekistan }


\graphicspath{{./figures}}

\makeatletter
\newcommand{\Jim}[1]{\textcolor{violet}{#1}}
\newcommand{\Rustam}{\@ifstar{\@Ra}{\@Rb}}
\newcommand{\@Rb}[1]{\textcolor[HTML]{0da34e}{#1}}
\newcommand{\@Ra}[1]{\textcolor[HTML]{0da34e}{\textbf{Rustam:} #1}}
\begin{abstract}
We present a single-shot near-field technique, where the near-field scan is performed on a single sample without repeating measurements or averaging over multiple samples, to reconstruct the isofrequency surfaces of metamaterials in the microwave regime. In our approach, we excite resonant modes using a fixed source in a resonator composed of the material under test and map the in-plane field distribution with a movable probe. Applying a fast Fourier transform (FFT) to the measured field reveals the sample’s in-plane dispersion. By extending this analysis over multiple frequencies and comparing the results with Fabry–Pérot resonances, we retrieve the full three-dimensional dispersion relation. When we apply the method to a double non-connected wire metamaterial, it accurately captures the low-frequency hyperbolic isofrequency surface, providing both a precise experimental tool and conceptual insight into spatially dispersive metamaterials.
\end{abstract}

\maketitle

\section{Introduction}

The recent development of metamaterials has significantly broadened the scope for exploring novel wave propagation phenomena. This progress stems from the unprecedented design flexibility offered by meta-atoms—the fundamental building blocks of metamaterials. Metamaterials have enabled experimental demonstrations of negative refraction \cite{shalaev2007optical}, backward waves \cite{silin2012history}, and other unconventional electromagnetic effects \cite{capolino2017theory}. Additionally, recent designs have exhibited exceptionally high effective permittivity along specific propagation directions \cite{poddubny2013hyperbolic, enriquez_2024}, surpassing the intrinsic limits of natural materials such as ceramics \cite{kell1973high}, oxides \cite{taylor2019fundamental,peng2020origin}, perovskites \cite{peng2020new}, and metal–dielectric composites in the low-frequency regime \cite{du2016colossal}. In particular, hyperbolic metamaterials \cite{poddubny2013hyperbolic}—a class of anisotropic media with permittivity tensor components of opposite sign—exhibit hyperbolic isofrequency contours. These materials support extreme electromagnetic responses across a wide frequency range, from microwaves to the optical domain.

The concept of dispersion provides a fundamental framework for describing wave phenomena in both continuous media and structured materials. Dispersion relations characterize how wave behavior depends on frequency and wave vector via the material’s effective parameters. Temporal dispersion refers to the frequency dependence of these parameters, while spatial dispersion accounts for nonlocal effects, in which the material response explicitly depends on the wave vector \cite{landau1960course, born}. Analyzing a material’s dispersion relations offers valuable insight into the interaction of both propagating and evanescent waves with the structure.

Dispersion diagrams, isofrequency contours and surfaces have long been widely accepted as methods for describing and analyzing the electromagnetic properties of materials and wave propagation \cite{joannopoulos2008molding, tretyakov2003analytical}.
Analytical determination of these dispersion characteristics necessitates knowledge of the material parameter tensors \cite{born}. For  metamaterials, which are the focus of the present study,  various homogenization methods can be used to determine the effective material parameters \cite{alu2011first,silveirinha2007metamaterial}. However,  such estimates  often fail to be universal over a wide frequency range.

\begin{figure*}[t]
    \center
    \includegraphics[width=1\linewidth]{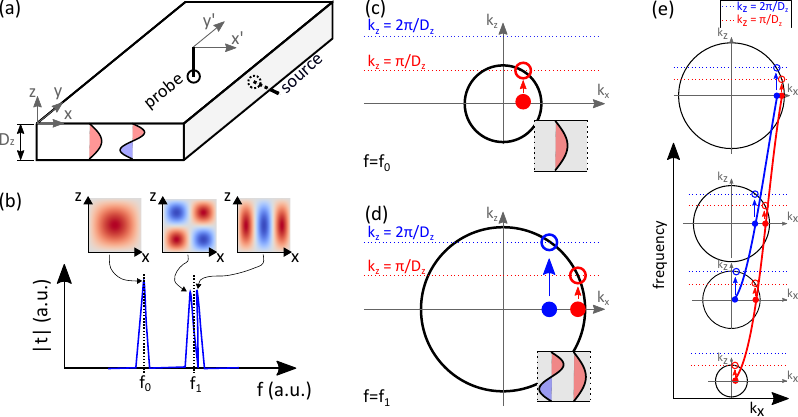}
    \caption{(a) Field-scanning setup with a fixed source and a movable probe that scans the field in the $x$–$y$ plane above the sample.
(b) Each resonance corresponds to a peak in the transmission spectrum. Applying a FFT to the scanned field yields the $k_x$ dependence of the field, assuming a fixed $k_y$ (i.e., slab waveguide with $k_y = 0$).
(c) At low frequencies ($f = f_0$), only the fundamental mode along $z$ is supported, characterized by $k_z = n_z \pi / D_z$ with $n_z = 1$.
(d) As the frequency increases ($f = f_1$), a higher-order mode ($n_z = 2$) appears.
(e) Tracking the guided modes across several frequencies, e.g., the first two guided modes, provides discrete points in constant frequency cuts of the dispersion relation $\omega(k_x, k_z)$.}
    \label{idea}
\end{figure*}

Developing a general-purpose algorithm for the theoretical description of arbitrary metamaterials is practically impossible due to their variety and the limitations of existing homogenization approaches. On the other hand, the dispersion of a specific metamaterial can often be readily calculated by solving Maxwell’s equations numerically \cite{capolino2017theory}. In many cases, it is sufficient to know the unit cell geometry and the constituent materials, without delving into all the physical subtleties. However, while experimental prototypes have been fabricated \cite{poddubny2013hyperbolic}, experimentally verifying the dispersion properties predicted by theory or simulations remains a challenging task.




Researchers have studied the dispersion properties of planar structures in the optical regime using back focal plane microscopy, which enables reconstruction of isofrequency contours \cite{permyakov_visualization_2018, pidgayko_direct_2019, jin_topologically_2019}. Isofrequency contours have also been obtained from numerical simulations of infinite cavity samples of different sizes, each resonating at the same frequency \cite{yang2012experimental}. In the microwave range, several techniques have been used to reconstruct dispersion characteristics of structured materials, including coherent microwave transient spectroscopy \cite{robertson_measurement_1992, tang_experimental_2023}, near-field scanning combined with a fast Fourier transform (FFT) \cite{lemoult_resonant_2010, yves_measuring_2018, tremain2018isotropic, hooper2014massively}, and complex-field measurements, in which wave vectors are inferred from phase differences per unit length \cite{belov2013single, itoh2005electromagnetic}. However, full three-dimensional dispersion properties—including isofrequency surfaces—have not yet been retrieved in a single measurement without rotating the sample.


Our study proposes an experimental approach that requires a single measurement of the electromagnetic near-field distribution of a multi-mode resonator. This method allows for comprehensive three-dimensional dispersion characterization across a wide frequency range by extracting isofrequency contours of several waveguide modes in the $k_xk_y$-plane. The $k_z$ component of each mode is identified  via the Fabry-Pérot resonance condition along the $z$-direction. We showcase the effectiveness of our approach through an example wherein we reconstruct the three-dimensional isofrequency surfaces of the hyperbolic mode within the double non-connected wire metamaterial \cite{simovski2004low}.


\section{Main concept}



Figure~\ref{idea}(a) presents a schematic of the proposed three-dimensional dispersion extraction method. A source excites waves that propagate through the sample, reflect from its boundaries, and interfere within the enclosed volume. A movable probe scans the top $x$–$y$ plane, measuring the electromagnetic field distribution point by point. While the in-plane dispersion relation $\omega(k_x, k_y)$ can be obtained using a standard FFT technique~\cite{tremain2018isotropic}, the out-of-plane wave-vector component $k_z$ is not known \textit{a priori}. However, when the refractive index of the sample is much higher than that of air—or when the sample is bounded by metallic walls—the system supports resonant modes that satisfy the Fabry–Pérot condition $D_i k_i = n_i \pi$, where $i = x, y, z$, $D_i$ is the sample dimension along the $i$ direction, and $n_i$ is an integer representing the resonance order. This relation enables discrete sampling of the dispersion along $k_z$, since each spectral peak---where $k_x$ and $k_y$ are already determined---corresponds to a resonant mode with a specific $n_z$ value.

To illustrate the underlying mechanism, we consider an isotropic high-index slab of thickness $D_z$ along the $z$ axis, surrounded by air, where $H_z$ waves are excited by a source and detected by a movable probe [see Fig.~\ref{idea}(a)]. For simplicity, we restrict the analysis to $k_y = 0$, corresponding to a slab waveguide that enables direct visualization of constant-frequency cuts of the dispersion relation $\omega(k_x, k_z)$. The transmission spectrum between the source and the probe exhibits discrete peaks associated with the cavity resonances [Fig.~\ref{idea}(b)]. At low frequencies [Fig.~\ref{idea}(c)], only the fundamental mode is present, with $k_z = \pi / D_z$, whereas at higher frequencies additional modes appear, such as the second-order mode with $k_z = 2\pi / D_z$ [Fig.~\ref{idea}(d)]. In general, the resonant modes satisfy $k_z = n_z \pi / D_z$, where $n_z$ denotes the mode order. The FFT of the measured field provides the in-plane wavevector $k_x$ for each resonance, which, together with the Fabry–Pérot condition along the $z$-axis, defines a point on the isofrequency contour. In the isotropic case, these contours are circular [Fig.~\ref{idea}(d)]. By repeating the measurement over a range of frequencies, one obtains the full spectrum $H_z(k_x, f)$, in which each dispersion branch corresponds to a distinct $k_z$ value [Fig.~\ref{idea}(e)], allowing reconstruction of the two-dimensional dispersion surface in the $k_xk_z$ plane.

To extend this approach to a slab of finite thickness $D_z$, where the waveguide modes exhibit field variations along both the $x$ and $y$ directions and are associated with different $k_z$ values, we aim to isolate the constant-frequency cuts of the dispersion relation $\omega(k_x, k_y)$ corresponding to each mode indexed by $n_z$. To visualize the contribution of multiple modes with distinct $n_z$ values, we analyze a series of cross-sections $H_z(k_x, f)$ at fixed $k_y$, each containing the relevant dispersion branches. By collecting, interpolating, and assigning these branches to the appropriate $k_z$ values consistent with the Fabry-Pérot resonances, we construct a four-dimensional matrix $H_z(k_x, k_y, k_z, f)$. This dataset can then be represented as a set of isofrequency contours $H_z(k_x, k_y)$ for given $k_z$ and $f$.

\section{Materials and Methods}
\subsection{Wire metamaterials}

To illustrate the method in detail, we consider an artificial material composed of metallic wires. Among all metamaterials, the class of \textit{wire media} is particularly notable for its pronounced spatial dispersion (nonlocality), which arises from the unrestricted flow of charges along the metallic wires. Although various types of wire-based metamaterials have been proposed, only specific subclasses have been rigorously described using an effective permittivity tensor. The \textit{simple wire medium}~\cite{belov2003strong}, as well as double and triple non-connected wire media~\cite{simovski2004low, sakhno2022quadraxial}, belong to this category. In contrast, other wire metamaterials still lack a comprehensive analytical description.

The method of isofrequency surface reconstruction introduced here provides a powerful tool to bridge this gap. It serves a dual purpose: validating existing effective-medium models and corroborating numerical simulations of nonhomogenized metamaterials. In this work, we focus on the double non-connected wire medium, which exhibits a hyperbolic-like mode at low frequencies~\cite{simovski2004low}.

\subsection{Double non-connected wire metamaterial}
\begin{figure}[h!]
    \center{\includegraphics[width=1\linewidth]{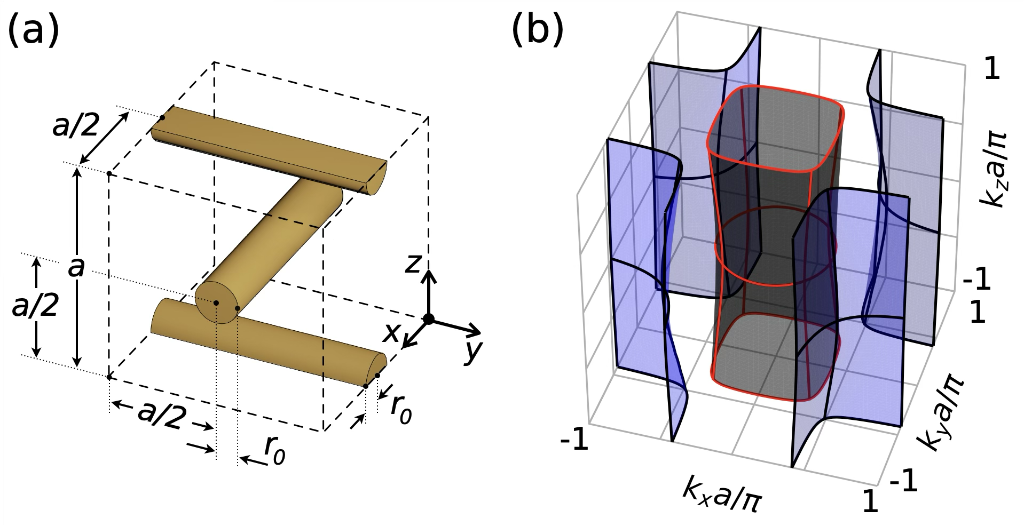}}
    \caption{(a) Unit cell of the double non-connected wire metamaterial. (b) Isofrequency surface $\omega=0.2 (2\pi c/a)$ (below the plasma frequency) plotted according to the effective medium model of the metamaterial} \cite{simovski2004low}.
    \label{fig:01_uc}
\end{figure}

The double non-connected wire metamaterial consists of two arrays of parallel, infinitely long wires with identical cross-sections. In one array, the wires are aligned along the $x$-axis, while in the other they are aligned along the $y$-axis. Each array is arranged in a square lattice with lattice constant $a$. The distance between two nearest perpendicular wires (specifically, the closest pair from different arrays) is $a/2$. A cubic unit cell of the metamaterial is depicted in Fig.~\ref{fig:01_uc}(a).

Solving the system of equations (40) from \cite{simovski2004low} with the effective permittivity tensor of the metamaterial substituted (see Eqs. (42–43) of \cite{simovski2004low}) leads to Eq. (36) of the same reference. At a fixed frequency, the determinant of the system (40) vanishes (which is equivalent to satisfying Eq. (36)) for the eigen wave vectors of the metamaterial. Consequently, the isofrequency surfaces can be obtained and plotted in reciprocal space. In Fig. \ref{fig:01_uc}(b), we present analytical isofrequency surfaces for the \textit{homogenized} \cite{belov2003strong, silveirinha2004hybrid, silveirinha2007metamaterial} double non-connected wire metamaterial below its plasma frequency.


The first mode (black surface, Fig.~\ref{fig:01_uc}(b)) is an \textit{ordinary mode}, also known as the low-$k$ mode~\cite{simovski2004low}, with TE polarization, where the electric field points along the $z$-axis [see Fig.~\ref{fig:01_uc}(a)]. In the $k_z=0$ plane, its isofrequency surface forms a circular cross-section with a radius equal to the vacuum wave vector at the corresponding frequency. In cross-sections perpendicular to $k_z$, the in-plane wave vector projection remains close to the vacuum wave number, reflecting the negligible polarizability of the thin wires.

The second mode (blue surface, Fig.~\ref{fig:01_uc}(b)) is an \textit{extraordinary mode}, featuring a hyperbolic-like isofrequency contour in the $k_xk_y$ plane (high-$k$ mode). It supports large wave vectors over a wide frequency range, up to the edge of the Brillouin zone, limited only by the structure’s periodicity. This mode exhibits TM polarization, producing a non-zero magnetic field component along the $z$-axis [see Fig.~\ref{fig:01_uc}(a)]. 

In our experiments, TM modes in the double non-connected wire medium exhibit standing-wave patterns in the $xy$ plane resembling conventional cavity modes. Although the effective permittivity along the wires is negative below the plasma frequency, the finite sample size provides strong reflections at the boundaries, effectively forming a cavity. The magnetic field along $z$ can still form discrete resonances along all three axes, approximately satisfying Fabry--Pérot conditions, with integers $n_x$, $n_y$, and $n_z$ denoting the number of field variations along $x$, $y$, and $z$, respectively.

\subsection{Multi-mode Resonator}
\begin{figure}[h!]
    \center{\includegraphics[width=1\linewidth]{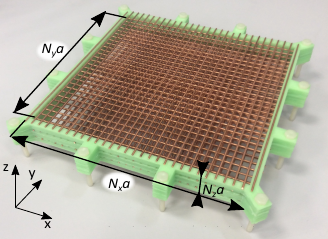}}
    \caption{Double non-connected wire media multi-mode resonator: a brick consisting of $N_x \times N_y \times N_z$ unit cells, with $N_x = 30$, $N_y = 30$, and $N_z = 3$. The metamaterial period is $a = 5.7~\mathrm{mm}$, and the wire radius is $r = 0.6~\mathrm{mm}$. The wire holder was 3D printed from ABS plastic (dielectric constant $\varepsilon' = 2.4$, loss tangent $\tan \delta \approx 0.01$, 100\% fill).
}
    \label{fig:02_sample}
\end{figure}

Our experimental sample consists of a multi-mode resonator filled with a double non-connected wire medium. A photograph of the prototype is shown in Fig.~\ref{fig:02_sample}. Copper wires (radius $r = 0.6$ mm) are held in place along the perimeter by an ABS plastic holder ($\varepsilon_{\text{ABS}} = 2.4$, $\tan\delta \approx 0.01$ in the microwave frequency range). The lattice constant is $a = 5.7$ mm. The geometric parameters of the double non-connected wire medium were chosen so that the frequency range of interest, from 1 to 14 GHz, lies below the plasma frequency, allowing the study of the corresponding isofrequency surfaces. The resonator dimensions are $N_x a \times N_y a \times N_z a$, where $N_x = 30$, $N_y = 30$, and $N_z = 3$.

\subsection{Near-field scanning and FFT}
\begin{figure}[h!]
    \center{\includegraphics[width=1\linewidth]{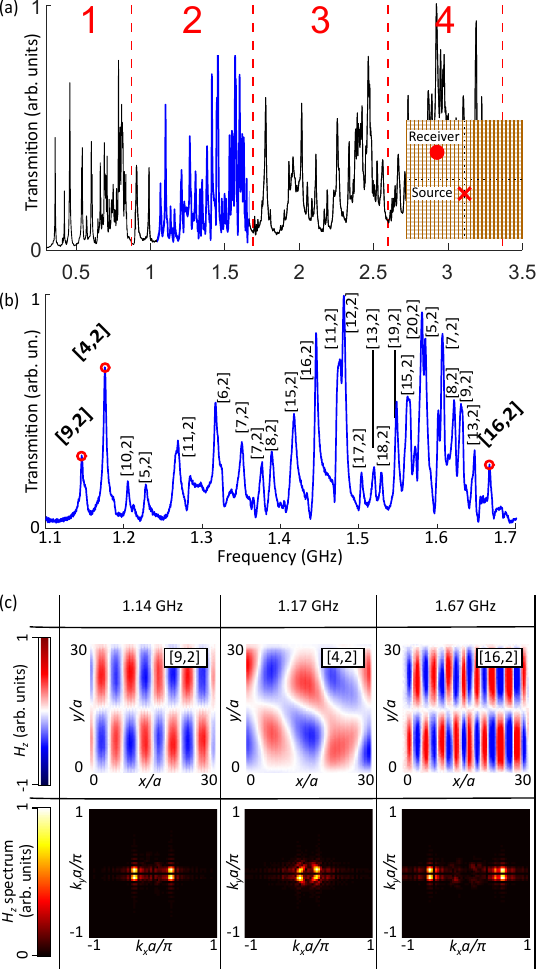}}
    \caption{Measurements of the $H_z$ field near the sample.
(a) Transmission coefficient between the source located and an arbitrarily positioned receiver; the numbers indicate the number of half-wavelengths across the wire array.
(b) Frequency range covering all responses with the same $k_y = 2\pi / (N_y a)$ order.
(c) Examples of near-field distributions in real space for selected frequencies (red dots in panel b), along with the corresponding Fourier spectra.}
    \label{fig:03_measurements}
\end{figure}

\begin{figure*}[t]
    \center{\includegraphics[width=0.9\linewidth]{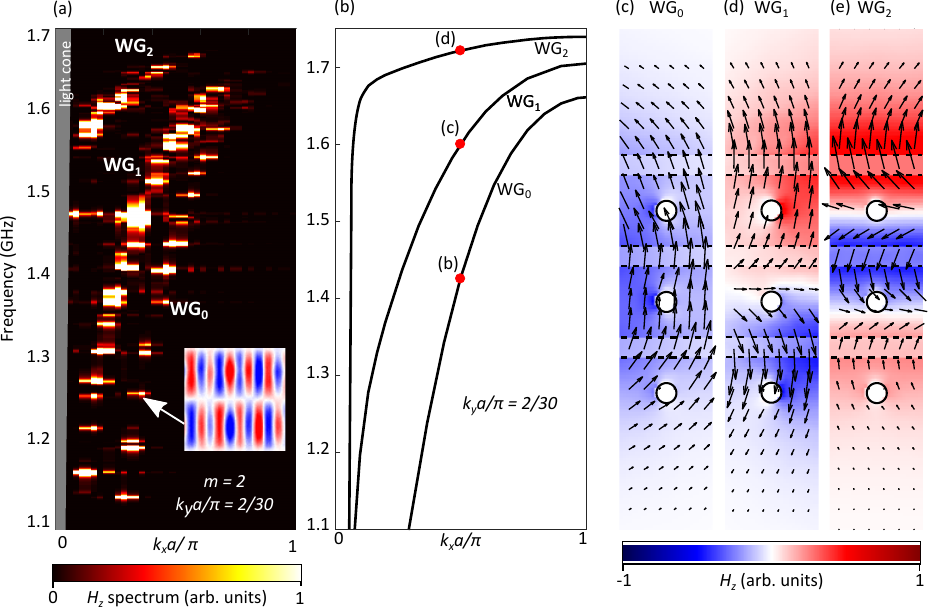}}
    \caption{Dispersion branches of different waveguide modes.  
(a) Experimentally obtained Fourier spectrum showing frequency versus $k_x$ for $k_y a / \pi = 2/30$.  
(b) Numerically calculated dispersion relation for the same $k_y$, obtained by solving the eigenmode problem.  
(c–e) Numerically calculated $H$-field distributions of the corresponding waveguide modes, with arrows indicating the magnetic field in the $x$–$z$ plane.}
   
    \label{fig:04_branches}
\end{figure*}

In our experimental setup, the near-field distribution above the sample was measured. Owing to the polarization of the high-$k$ modes, the measured transmission parameter S$_{21}$ is proportional to the $H_z$ field distribution at the corresponding frequency. Using a vector network analyzer (VNA), we recorded the transmission between two electrically small loop antennas over a broad frequency range [see Fig.~\ref{fig:03_measurements}(a,b)]. The source antenna was placed near the center of the resonator, inside the sample, to excite all supported TM modes through near-field coupling. The receiving antenna—also a small loop oriented to detect the magnetic field component normal to the sample surface—was positioned 3 mm above the sample. For the spectra presented in Fig.~\ref{fig:03_measurements}(a,b), the receiver was placed at a position offset by fifteen unit cells in both the vertical and horizontal directions from the center of the sample, as indicated in the top-view inset of Fig.~\ref{fig:03_measurements}(a).



To perform eigenmode analysis of the experimental data, we employed a FFT, which converts the field distribution $H_z(x,y)$ in real space into the Fourier spectrum $H_z(k_x,k_y)$ in reciprocal space, also referred to as $k$-space. The resulting two-dimensional map has a limited resolution due to the finite number of unit cells in the resonator. Results for several frequencies are shown in Fig.~\ref{fig:03_measurements}(c), where the brightest points in $k$-space correspond to the wavevectors of the plane-wave components with the highest amplitude in real space, forming the isofrequency contours.

Thus, we construct a three-dimensional intensity matrix $H_z(k_x, k_y, \omega)$ by \textit{stacking} FFT-generated Fourier spectra at different frequencies. From this matrix, it is possible to select a specific slice $H_z$ corresponding to a fixed $k_y = n_y \pi / (N_ya)$, which satisfies the $n_y$-th order Fabry–Pérot resonance condition.

Although the probe-antenna method is inherently limited to the microwave regime, the underlying approach—reconstructing isofrequency surfaces from spatially resolved field measurements—is conceptually applicable to higher-frequency domains; in optics, for example, near-field scanning probes and Fourier/back-focal-plane or leakage-radiation imaging have been used to obtain $k$-space (isofrequency) maps and extract dispersion of guided and surface modes \cite{Betzig1992, Regan2016, Vasista2019, Jang2024}.

\section{Results and discussion}

\begin{figure*}[t]
    \center{\includegraphics[width=0.95\textwidth]{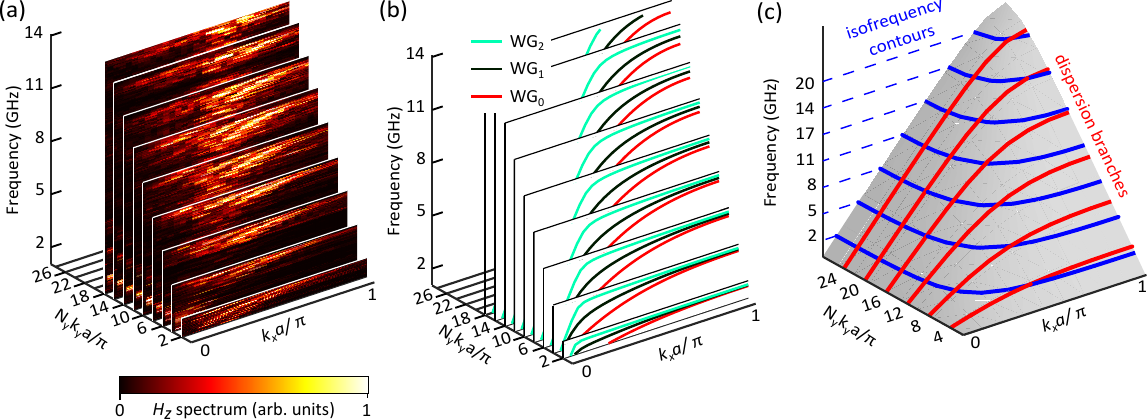}}
    \caption{Conversion of measured Fourier spectra into isofrequency contours.  
(a) Experimentally obtained Fourier spectra for $N_y k_y a / \pi = 2, 4, \dots, N_y$ with $N_y = 30$ unit cells of the metamaterial along $y$, showing three dispersion branches.  
(b) Corresponding dispersion branches extracted from the Fourier spectra.  
(c) Dispersion branches of the fundamental waveguide mode ($\mathrm{WG_0}$) mapped into isofrequency contours.}
    \label{fig:05_contours}
\end{figure*}
\begin{figure*}[t]
    \center{\includegraphics[width=0.9\linewidth]{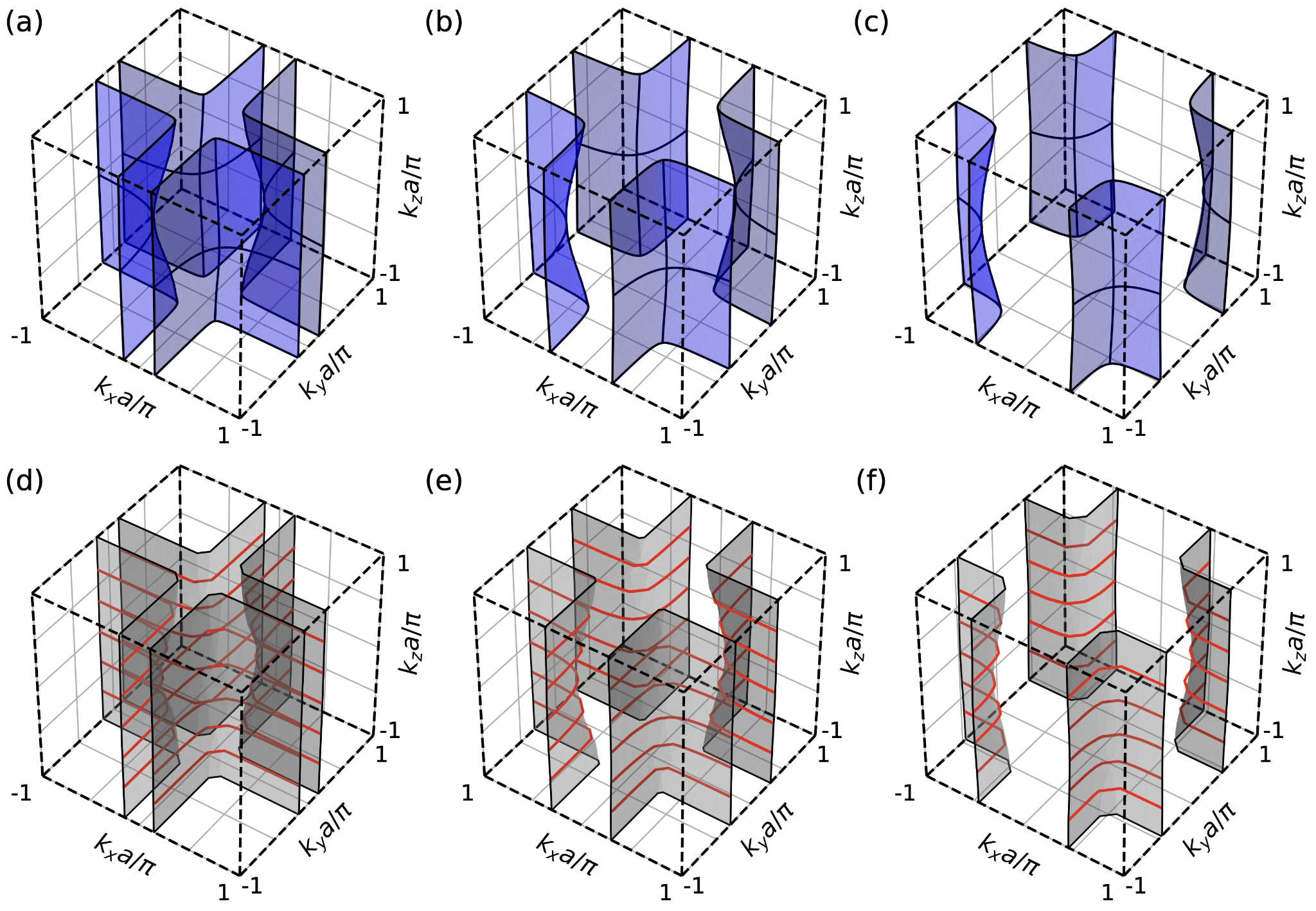}}
    \caption{(a–c) Analytically obtained isofrequency surfaces of the double non-connected wire medium.  
(d–f) Experimentally measured isofrequency contours (red lines) plotted alongside numerically calculated surfaces (gray).  
Frequencies shown (below the plasma frequency): (a, d) 4 GHz; (b, e) 8 GHz; (c, f) 12 GHz.}
    \label{fig:06_disp}
\end{figure*}

The experimentally obtained $k_y$-plane slice of the matrix $H_z(k_x, k_y, \omega)$ appears in Fig.~\ref{fig:04_branches}(a). The cross-section displays the dispersion of individual waveguide modes as a function of frequency versus $k_x$ for fixed $k_y$, $k_ya/\pi=2/30$. Three distinct curves emerge, each composed of discrete resonance points. 

To investigate the properties of the waveguide modes, we performed eigenmode simulations using CST STUDIO SUITE \cite{cst}. The numerical model implemented periodic boundary conditions in the wire plane ($x$-$y$), and absorbing boundary conditions (or Perfectly Matched Layers (PMLs)) were applied in the $z$ direction. The $k_y$ component was fixed to the experimentally determined value [see Fig.~\ref{fig:04_branches}(a)], while the $k_x$ wave vector was swept across the first Brillouin zone, ranging from $0$ to $\pi/a\,\mathrm{m}^{-1}$. The resulting dispersion diagram, shown in Fig.~\ref{fig:04_branches}(b), reveals three distinct branches: the fundamental mode, $\mathrm{WG}_0$ [Fig.~\ref{fig:04_branches}(c)], and a pair of high-order modes [Figs.~\ref{fig:04_branches}(d) and \ref{fig:04_branches}(e)].

The analysis of the magnetic field (shown in arrows) distributions confirms that the polarization is commonly TM, and the waveguide modes are characterized by the number of half-wavelengths hosted inside across the layer, namely $0$, $1$ and $2$. The extraction of the dispersion branches facilitates the individual analysis of the properties of each waveguide mode.
Sampling of wavevector components due to Fabry–Pérot resonances within the finite resonator separates waveguide modes with different $k_z$ components. The number of such modes corresponds to the structure’s thickness in terms of unit cells, yielding three waveguide modes, as shown in Fig.~\ref{fig:04_branches}(c). Each resonance curve depicts the dispersion of a mode with a specific order $n_z = 0, 1, 2$.



To extract the dispersion branches from a specific $k_y$-slice of the spectrum map [Fig.~\ref{fig:05_contours}(a)], we manually select the three sets of bright points and then apply a \textit{least-squares interpolation} to generate smooth dispersion curves, as shown in Fig.~\ref{fig:05_contours}(b). Interestingly, researchers in geophysics face a similar challenge when separating dispersion branches and have addressed it using deep learning methods~\cite{dong2021dispernet, dai2021deep}.

Using the above technique, all possible dispersion branches were sequentially collected for the different $k_z$ values (waveguide modes $WG_i$, with $i = 0, 1, 2$), as shown in Fig.~\ref{fig:05_contours}(b). Two-dimensional dispersion diagrams were then derived separately for each waveguide mode. Dispersion branches and isofrequency contours can be obtained from cross-sections of the $f(k_x, k_y)$ surface along different planes, establishing a direct connection between them, as illustrated in Fig.~\ref{fig:05_contours}(c).

We reconstruct the isofrequency surfaces from the extracted two-dimensional dispersion $f(k_x, k_y)$ [Fig.~\ref{fig:05_contours}(c)] for each mode corresponding to $k_z = n_z \pi / (a N_z)$. The surfaces for selected frequencies below the plasma frequency of the wire-medium sample ($f = 4, 8, 12~\mathrm{GHz}$) are shown in Fig.~\ref{fig:06_disp}. Figures~\ref{fig:06_disp}(a–c) present the analytically obtained isofrequency surfaces~\cite{simovski2004low}, while Figs.~\ref{fig:06_disp}(d–f) display the corresponding experimentally measured contours (red lines) together with the numerical results obtained from the CST eigenfrequency solver~\cite{cst} (gray surfaces). The experimentally reconstructed contours show excellent agreement with both the analytical and numerical results, reproducing the hyperbolic behavior in the $k_xk_y$ plane expected for the TM mode. 

Increasing the number of unit cells within the resonator improves the reconstruction accuracy by introducing additional Fabry–Pérot resonances within the frequency range. However, experimental limitations can still lead to incomplete reconstruction of the isofrequency surfaces. First, the amplitude of each resonance strongly depends on the coupling between the antenna and the sample’s eigenmodes; as a result, not all resonances are observable. Furthermore, the three-dimensional nature of the modes can cause resonant frequencies corresponding to different $k_z$ values to overlap, reducing the number of distinct samples available for reconstructing the dispersion along the $k_z$ direction.

When the refractive index of the sample ($n_2$) is not much higher than that of air ($n_1$), the Fabry--Pérot model no longer provides an accurate approximation of the dispersion along $z$. More accurate results can be obtained using methods from dielectric waveguide theory; however, the range over which isofrequency contours can be constructed becomes limited. For waves confined in two transverse directions, the transverse wavevector components $k_x$ and $k_y$ are quantized by the boundary conditions, and the allowed propagation constants satisfy $k_0 n_1 < k_z < k_0 n_2$. As a result, $k_z$ is restricted to a narrow interval for a given frequency, reducing the accessible portion of the dispersion surface.


\section{Conclusion}

We have introduced a method to experimentally reconstruct isofrequency surfaces in three-dimensional metamaterials. The method relies on a single-shot near-field scan of the resonant modes of a resonator composed of the target metamaterial. By applying fast Fourier transforms over a wide frequency range and comparing the observed resonances with a Fabry–Pérot model, we obtain the isofrequency surfaces. We have demonstrated the method’s effectiveness using a resonator based on a double non-connected wire metamaterial, confirming the hyperbolic isofrequency surface of one of its two low-frequency modes. The approach is straightforward to implement and enables rapid, comprehensive dispersion characterization. Although it requires samples with a refractive index significantly higher than that of air, it can be extended to more general configurations where electric walls bound the structure; in such cases, field scans can be performed through predrilled holes in the cavity, as shown in Ref.~\cite{enriquez_2025}.


\begin{acknowledgments}
The work is funded by Russian Science Foundation grant No. 25-12-00261 (\href{https://rscf.ru/project/25-12-00261/}{https://rscf.ru/project/25-12-00261/}).
\end{acknowledgments}
\section*{Data availability}
The data that support the findings of this article are
openly available \cite{koreshin_2026_18503520}.

\bibliography{bibliography} 

\end{document}